\documentclass[11pt,a4paper]{article}
\usepackage[T1]{fontenc}
\usepackage[utf8]{inputenc}
\usepackage{lmodern}
\usepackage[margin=2.55cm,top=2.55cm,bottom=2.55cm]{geometry}
\usepackage{amsmath,amssymb,amsthm}
\usepackage{booktabs}
\usepackage{graphicx}
\usepackage{float}
\usepackage{microtype}
\usepackage{comment}
\usepackage{tikz}
\usetikzlibrary{arrows.meta,calc,positioning,fit,backgrounds,matrix,decorations.pathreplacing}
\usepackage[hidelinks]{hyperref}
\hypersetup{
  pdftitle={Reopening the one clean qubit case of a semi-grand challenge},
  pdfauthor={Christopher W. Mansell},
  bookmarksnumbered=true
}
\theoremstyle{definition}
\newtheorem{theorem}{Theorem}
\newtheorem{continuousproposition}{Proposition}[section]

\newcommand{\TraceCircuitFigure}{%
  \begin{figure}[htbp]
    \centering
    \begin{tikzpicture}[
      x=1cm,y=1cm,
      line width=.55pt,
      circuitgate/.style={draw,fill=white,minimum width=.48cm,
        minimum height=.43cm,inner sep=0pt},
      font=\normalsize
    ]
      \draw (0,.75) -- (3.60,.75);
      \draw (0,0) -- (3.60,0);
      \draw (.54,-.11) -- (.70,.11);
      \node[anchor=south west,inner sep=1pt,font=\scriptsize] at (.68,.08) {$n$};
      \node[anchor=east,inner sep=4pt] at (0,.75) {$|+\rangle$};
      \node[anchor=east,inner sep=4pt] at (0,0) {$\mathbb{I}/2^{n}$};
      \draw (1.49,.75) -- (1.49,0);
      \fill (1.49,.75) circle[radius=.050cm];
      \node[circuitgate] at (1.49,0) {$U$};
      \draw[fill=white] (2.98,.535) rectangle (3.60,.965);
      \draw (3.065,.665) arc[start angle=160,end angle=20,
        x radius=.238cm,y radius=.145cm];
      \draw[-{Stealth[length=1.5mm,width=1.0mm]}]
        (3.245,.640) -- (3.485,.900);
    \end{tikzpicture}
    \caption{The quantum circuit for estimating the normalized trace of a unitary matrix, $U$.
    The lower wire denotes an $n$-qubit maximally mixed register.
    Measuring the clean qubit in the $X$ basis gives the real part of the normalized trace, while measuring in the $Y$ basis gives the imaginary part.}
    \label{fig:trace}
  \end{figure}%
}

\newcommand{\CounterexampleCircuitFigure}{%
  \begin{figure}[htbp]
    \centering
    \begin{tikzpicture}[
      x=1cm,y=1cm,
      line width=.55pt,
      circuitgate/.style={draw,fill=white,minimum width=.37cm,
        minimum height=.37cm,inner sep=0pt},
      font=\normalsize
    ]
      \draw (0,1) -- (4.80,1);
      \draw (0,.5) -- (4.80,.5);
      \draw (0,0) -- (4.80,0);
      \node[anchor=east,inner sep=4pt] at (0,1) {$a$};
      \node[anchor=east,inner sep=4pt] at (0,.5) {$b$};
      \node[anchor=east,inner sep=4pt] at (0,0) {$c$};

      \draw (.80,1) -- (.80,.5);
      \fill (.80,1) circle[radius=.045cm];
      \draw[fill=white] (.80,.5) circle[radius=.10cm];
      \draw (.70,.5) -- (.90,.5);
      \draw (.80,.40) -- (.80,.60);

      \draw[gray,dashed,line width=.5pt]
        (1.67,-.26) rectangle (3.18,1.32);
      \draw (2.08,1) -- (2.08,0);
      \draw[fill=white] (2.08,1) circle[radius=.065cm];
      \node[circuitgate] at (2.08,0) {$Z$};
      \node[circuitgate] at (2.76,1) {$T$};

      \draw (4.07,.5) -- (4.07,0);
      \fill (4.07,.5) circle[radius=.045cm];
      \draw[fill=white] (4.07,0) circle[radius=.10cm];
      \draw (3.97,0) -- (4.17,0);
      \draw (4.07,-.10) -- (4.07,.10);

      \node at (.80,-.57) {$G_{1}$};
      \node at (2.425,-.57) {$G_{2}$};
      \node at (4.07,-.57) {$G_{3}$};
    \end{tikzpicture}
    \caption{The counterexample circuit on the register qubits $a$, $b$ and $c$.
      The gate $G_{2}$ (dashed box) applies $Z$ to $c$ when $a$ is $|0\rangle$
      (open circle) and the phase $e^{i \pi / 4}$ when $a$ is $|1\rangle$, where
      $T=\operatorname{diag}(1,e^{i \pi / 4})$.
      In the one clean qubit model, each gate is also controlled on the clean qubit that is not shown. 
      Hence, the complete circuit has four qubits.}
    \label{fig:counterexample}
  \end{figure}%
}

\newcommand{\BranchStructureFigure}{%
  \begin{figure}[htbp]
    \centering
    \begin{tikzpicture}[
      x=1cm,y=1.2cm,
      line width=.45pt,
      branchbox/.style={draw,rounded corners=1pt,minimum height=.48cm,
        minimum width=2.15cm,inner xsep=5pt,inner ysep=3pt,align=center},
      brancharrow/.style={-{Latex[length=1.8mm,width=1.25mm]}},
      font=\normalsize
    ]
      \node[branchbox,minimum width=1.75cm] (branch-root) at (0,2.80) {$C_{3}$};
      \node[branchbox,minimum width=1.75cm] (branch-zero) at (-3,1.40) {$C_{3,0}$};
      \node[branchbox,minimum width=2.25cm] (branch-one) at (3,1.40)
        {$C_{3,1}$};
      \node[branchbox] (branch-z) at (-4.50,0) {$Z_{c}$};
      \node[branchbox] (branch-xz) at (-1.50,0) {$(XZ)_{c}$};
      \node[branchbox] (branch-x) at (1.50,0) {$e^{i \pi / 4}X_{b}$};
      \node[branchbox] (branch-zx) at (4.50,0) {$e^{i \pi / 4}(ZX)_{b}$};

      \draw[brancharrow] (branch-root.south west) --
        node[pos=.51,above left,inner sep=3pt] {$a=0$} (branch-zero.north);
      \draw[brancharrow] (branch-root.south east) --
        node[pos=.51,above right,inner sep=3pt] {$a=1$} (branch-one.north);
      \draw[brancharrow] (branch-zero.south west) --
        node[pos=.50,above left,inner sep=3pt] {$b=0$} (branch-z.north);
      \draw[brancharrow] (branch-zero.south east) --
        node[pos=.50,above right,inner sep=3pt] {$b=1$} (branch-xz.north);
      \draw[brancharrow] (branch-one.south west) --
        node[pos=.50,above left,inner sep=3pt] {$c=+$} (branch-x.north);
      \draw[brancharrow] (branch-one.south east) --
        node[pos=.50,above right,inner sep=3pt] {$c=-$} (branch-zx.north);
    \end{tikzpicture}
    \caption{The unique branches of $C_{3}$.
      The first split is by the value of $a$ in the $Z$ basis.
      Within the branch $a=0$, the unitary $C_{3,0}$ branches according to the value of $b$ in the $Z$ basis.
      Within the branch $a=1$, the unitary $C_{3,1}$ branches according to the value of $c$ in the $X$ basis.
      The leaves show the single-qubit target gates.}
    \label{fig:branches}
  \end{figure}%
}

\newcommand{\CanonicalFormStructure}{%
  \begin{figure}[htbp]
    \centering
    \begin{tikzpicture}[
      x=1cm,y=1.05cm,
      line width=.45pt,
      branchbox/.style={draw,rounded corners=1pt,minimum height=.48cm,
        minimum width=2.15cm,inner xsep=5pt,inner ysep=3pt,align=center},
      brancharrow/.style={-{Latex[length=1.8mm,width=1.25mm]}},
      font=\normalsize
    ]
      \node[branchbox,minimum width=1.75cm] (branch-root) at (0,2.80) {qubit 1};
      \node[branchbox,minimum width=1.75cm] (branch-zero) at (-3.35,1.40) {qubit 2};
      \node[branchbox,minimum width=2.25cm] (branch-one) at (3.35,1.40)
        {qubit 2};
      \node[branchbox] (branch-z) at (-5.00,0) {$e^{i \theta_{00}} \bigotimes_{j=1}^{q}W^{j,00}_{T_{j}}$};
      \node[branchbox] (branch-xz) at (-1.75,0) {$e^{i \theta_{01}} \bigotimes_{j=1}^{q}W^{j,01}_{T_{j}}$};
      \node[branchbox] (branch-x) at (1.75,0) {$e^{i \theta_{10}} \bigotimes_{j=1}^{q}W^{j,10}_{T_{j}}$};
      \node[branchbox] (branch-zx) at (5.00,0) {$e^{i \theta_{11}} \bigotimes_{j=1}^{q}W^{j,11}_{T_{j}}$};

      \draw[brancharrow] (branch-root.south west) --
        node[pos=.51,above left,inner sep=3pt] {$x_{1}=0$} (branch-zero.north);
      \draw[brancharrow] (branch-root.south east) --
        node[pos=.51,above right,inner sep=3pt] {$x_{1}=1$} (branch-one.north);
      \draw[brancharrow] (branch-zero.south west) --
        node[pos=.50,above left,inner sep=3pt] {$x_{2}=0$} (branch-z.north);
      \draw[brancharrow] (branch-zero.south east) --
        node[pos=.50,above right,inner sep=3pt] {$x_{2}=1$} (branch-xz.north);
      \draw[brancharrow] (branch-one.south west) --
        node[pos=.50,above left,inner sep=3pt] {$x_{2}=0$} (branch-x.north);
      \draw[brancharrow] (branch-one.south east) --
        node[pos=.50,above right,inner sep=3pt] {$x_{2}=1$} (branch-zx.north);
    \end{tikzpicture}
    \caption{Branching structure of a unitary in canonical form, illustrated with two qubits in set $B$.
    Note that these are the same qubits (i.e., qubit 1 and qubit 2) throughout the branching structure.}
    \label{fig:canonical_form}
  \end{figure}%
}

\newcommand{\SimulationScopeFigure}{%
  \begin{figure}[ht]
    \centering
    \colorlet{scopeset}{black!55}
    \colorlet{scopesetfill}{black!7}
    \definecolor{scopepc}{RGB}{47,109,179}
    \colorlet{scopepcfill}{scopepc!18}
    \definecolor{scopecx}{RGB}{217,115,13}
    \begin{tikzpicture}[
      x=1cm,y=1cm,
      line width=.55pt,
      font=\small,
      chainbox/.style={draw=scopepc,fill=white,rounded corners=2pt,
        text width=5.6cm,minimum height=.85cm,align=flush center,inner sep=3pt},
      chainarrow/.style={-{Latex[length=1.8mm,width=1.25mm]},draw=scopepc},
      arrowlabel/.style={right,font=\footnotesize,align=left,inner sep=3pt},
    ]
      \node[font=\small\bfseries,anchor=north west] at (-4.45,3.85) {(a)};
      \filldraw[draw=scopeset,fill=scopesetfill] (-1,0) circle[radius=3.15cm];
      \filldraw[draw=scopepc,fill=scopepcfill] (0.45,-0.3) circle[radius=1.3cm];
      \node[anchor=south] at (-1,3.15) {$\mathrm{DQC1}_{\mathrm{sep}}$ circuits};
      \node[align=center,text=scopepc] at (0.45,-0.3) {Product\\control\\circuits};
      \node[align=center,font=\footnotesize\itshape,text=black!70] at (-1.05,1.8)
        {Claimed empty Thm.~4.3.9};
      \fill[scopecx] (-2.25,0.78) circle[radius=.09cm];
      \node[align=center,text=scopecx,anchor=north] at (-2.25,0.6)
        {Our\\counterexample\\(Fig.~\ref{fig:counterexample})};
      \node[align=center,text=scopecx,font=\footnotesize,anchor=north] at (-2.15,-0.75)
        {Final unitary\\has no canonical\\form (Thm.~\ref{theorem:noncanonical})};
      \node[font=\small\bfseries,anchor=north west] at (4.0,3.85) {(b)};
      \node[chainbox,text=scopepc] (pc)   at (6.75,2.6)  {Product control circuit};
      \node[chainbox]              (rep)  at (6.75,1.0)  {Compact canonical representation};
      \node[chainbox]              (samp) at (6.75,-0.6) {Eigenvalue sampling and\\trace estimation};
      \node[chainbox,draw=scopeset,dashed] (bpp) at (6.75,-2.45)
        {$\mathrm{DQC1}_{\mathrm{sep}}\subseteq\mathrm{BPP}$};
      \draw[chainarrow] (pc.south) -- node[arrowlabel,text=scopepc] {Thm.~4.3.11} (rep.north);
      \draw[chainarrow] (rep.south) -- node[arrowlabel,text=scopepc] {Thm.~4.3.12} (samp.north);
      \draw[chainarrow,draw=scopeset,dashed] (samp.south) --
        node[arrowlabel,text=black!70] {only with Thm.~4.3.9\\(claimed)} (bpp.north);
    \end{tikzpicture}
    \caption{The relationship between the simulation argument of Ref.~\cite{ref4} and the counterexample provided in the present work.
    All theorems except Theorem~\ref{theorem:noncanonical} are those of Ref.~\cite{ref4}.
    (a) The outer circle represents $\mathrm{DQC1}_{\mathrm{sep}}$ circuits, which, according to Theorems~4.2.4 and~4.2.5, are exactly the circuits where every accumulated unitary has a separable eigendecomposition.
    The inner circle represents product control circuits, and Theorem~4.3.9 claims that the region between the two circles is empty.
    Our counterexample circuit, shown in Figure~\ref{fig:counterexample}, lies in this region: it is a $\mathrm{DQC1}_{\mathrm{sep}}$ circuit whose third gate does not follow the specified rules (see Section~\ref{sec:false-necessity} for further details), and its final accumulated unitary has no canonical form (see Theorem~\ref{theorem:noncanonical}).
    (b) The proof of Theorem~4.3.11 contains an efficient classical algorithm for turning a product control circuit into a compact canonical representation.
    The classical algorithm described in the proof of Theorem~4.3.12 uses this representation to efficiently sample eigenvalues and estimate the normalized trace.
    In order for these classical algorithms to apply to all $\mathrm{DQC1}_{\mathrm{sep}}$ circuits, Theorem~4.3.9 needs to hold.
    }
    \label{fig:scope}
  \end{figure}%
}

\usepackage{authblk}
\newcommand*{\myat}{{\fontfamily{ptm}\selectfont @}}

\begin{document}

\title{Reopening the one clean qubit case of a semi-grand challenge}
\author{Christopher W. Mansell\thanks{{cm\myat terraquantum.swiss}}}
\affil{Terra Quantum UK Ltd., London, United Kingdom}
\date{\today}
\maketitle

\begin{abstract}
In his 2005 list of semi-grand challenges for quantum computing theory, Scott Aaronson asked whether a quantum computer that is always in a separable mixed state can be simulated by a classical computer in polynomial time.
An affirmative answer was claimed for the one clean qubit model of quantum computation.
The supporting mathematical arguments relied on an exhaustive characterization of circuits that stay in separable states.
Here, we give a counterexample where one pure qubit, known as the clean qubit, controls a sequence of three gates that act on a maximally mixed three-qubit register.
The overall unitary that accumulates as these gates are applied has the properties that ensure the state of all four qubits remains fully separable.
Crucially, we show that the final unitary cannot be put into the form required by the claimed simulation result.
We also give a continuous-time implementation of the same circuit that keeps the qubits fully separable throughout and ends at the same unitary.
We extend our counterexamples to a uniform family, and pinpoint the gap in the earlier proof.
Since our circuits are themselves efficient to simulate classically, this work does not settle Aaronson's question.
It does, however, reopen the semi-grand challenge for an actively researched model of quantum computation.
\end{abstract}

\section{Introduction}
\label{sec:introduction}

\paragraph{The challenge.} In 2005, Aaronson asked whether a quantum computer that is always in a separable mixed state can be simulated by a classical computer in polynomial time~\cite{ref1}.
The question asks whether a restriction on the states visited during a computation is enough to guarantee an efficient classical account of its observable output.
The one clean qubit model of quantum computation provides an intriguing setting in which to investigate this question.

\paragraph{The model.} Introduced by Knill and Laflamme~\cite{ref2}, the one clean qubit model of quantum computation involves just one pure qubit, known as the clean qubit, and $n$ maximally mixed qubits (see Figure~\ref{fig:trace}).
The complexity class associated with this model is known as deterministic quantum computation with one qubit (or DQC1 for short).
Despite the lack of purity, this model is remarkably powerful~\cite[Chapter~22]{ref100}.
Estimating the normalized trace of a unitary matrix is a DQC1-complete task for which no polynomial-time classical algorithm is likely to exist~\cite{ref101}, and related sampling tasks are classically hard under plausible complexity-theoretic assumptions~\cite{ref11,ref12,ref13}.
We use the trace-estimation formulation throughout.

\TraceCircuitFigure

\paragraph{What the circuit estimates.} In the trace-estimation formulation, a Hadamard gate prepares the clean qubit in $|+\rangle$.
It then controls an efficiently implementable register unitary $U$, as in Figure~\ref{fig:trace}.
The expectation values of $X$ and $Y$ on the clean qubit are, respectively, the real and imaginary parts of the normalized trace $\tau(U)=2^{-n}\operatorname{Tr}[U]$, which is the average of the eigenvalues of $U$.
Throughout, $n$ denotes the number of register qubits, and the operations upon them are always controlled by the clean qubit.
Appendix~\ref{app:background} gives the readout formulas and their derivation.

\paragraph{The importance.} At first glance, estimating the trace of an exponentially large unitary matrix does not seem like it has many applications.
However, normalized trace estimation connects DQC1 to knot and 3-manifold invariants~\cite{ref5,JordanWocjan2009,JordanAlagic2014}, partition functions~\cite{ref7}, infinite-temperature out-of-time-order and multi-time correlation functions~\cite{ref8}, topological data analysis~\cite{LoweKBH2026,GCD2022,CadeCrichigno2024}, Schatten norms~\cite{ref6}, variational quantum compiling~\cite{Khatri2019}, and restricted quantum computer access~\cite{ref24}, to name a few.

DQC1 is also of fundamental interest, both as an intermediate model in complexity theory~\cite{ref10,JacobsMehraban2024} and as a testbed for non-classical correlations and resource theories~\cite{DFC2005,ref17,Matera2016,Wang2019}.
Experimentally, the need for every operation to be controlled by the clean qubit appears daunting.
Despite this, neutral-atom systems have demonstrated high-fidelity three-qubit gates~\cite{ref26}, which supports proposals in which a single atom natively controls a large register~\cite{ref25}.
This strengthens the prospects for a sizeable implementation of this model.

\paragraph{The claimed resolution.} We now turn to the work of Yoganathan and Cade~\cite{ref3}, and its revised treatment in Chapter~4 of Yoganathan's thesis~\cite{ref4}.
These works aimed to provide an exhaustive characterization of fully separable one clean qubit circuits and argued that the eigenvalues of their final unitaries could be efficiently sampled.
They thus claimed that one clean qubit circuits that are always in separable mixed states can indeed be simulated by classical computers in polynomial time.

\paragraph{Our counterexample.} In Section~\ref{sec:counterexample}, we analyze the quantum circuit in Figure~\ref{fig:counterexample} and show that it is fully separable after each discrete gate.
In Section~\ref{sec:noncanonical}, we prove that the final register unitary cannot be put in the required canonical form.
The resulting structural obstruction invalidates the proposed simulation argument and reopens the one clean qubit case of the semi-grand challenge.

\CounterexampleCircuitFigure

\paragraph{Organization.} Section~\ref{sec:separability} gives the properties of the gates that ensure the state of the circuit is always fully separable.
Section~\ref{sec:counterexample} applies this criterion to our counterexample.
Section~\ref{sec:canonical} introduces the canonical form of fully separable circuits and Section~\ref{sec:noncanonical} proves that our counterexample cannot be put into this form.
Section~\ref{sec:uniform} gives a uniform family of counterexamples and describes the consequences for classical simulations of one clean qubit circuits.
Finally, Section~\ref{sec:conclusion} provides a conclusion and an outlook.

\paragraph{Conventions.} Formally, $\mathrm{DQC1}_{\mathrm{sep}}$ is the class of decision problems that can be efficiently decided in the one clean qubit model with no entanglement across any cut during the computation.
This terminology may be used informally (e.g., ``$\mathrm{DQC1}_{\mathrm{sep}}$ circuits'') where there is no risk of misunderstandings arising.
We employ the gate convention of Refs.~\cite{ref3,ref4} where each displayed one- or two-qubit gate is additionally controlled by the clean qubit.
By efficient classical simulation, we mean polynomial-time estimation of the final clean-qubit measurement probabilities to inverse-polynomial additive accuracy, as in the cited work.
Here, polynomial is with respect to the number of qubits.
The decision-class inclusion claimed in Refs.~\cite{ref3,ref4} is $\mathrm{DQC1}_{\mathrm{sep}}\subseteq\mathrm{BPP}$, where $\mathrm{BPP}$ denotes efficient randomized classical decision computation.

Depending on context, we may use $|x_{\pm}\rangle = |\pm\rangle =(|0\rangle \pm |1\rangle)/\sqrt{2}$ and $|y_{\pm}\rangle=(|0\rangle \pm i|1\rangle)/\sqrt{2}$ to respectively denote the $\pm1$ eigenvectors of the Pauli operators $X$ and $Y$.
Similarly, $|z_{+}\rangle=|0\rangle$ and $|z_{-}\rangle=|1\rangle$ are the eigenvectors of the Pauli $Z$ operator.

\section{Separability in \texorpdfstring{$\mathrm{DQC1}_{\mathrm{sep}}$}{DQC1sep} circuits}
\label{sec:separability}

\paragraph{Clarifying separability.} A mixed state is separable across a bipartition if it can be written as a probabilistic mixture of pure states that are tensor products across that bipartition~\cite[Definition~4.2.1]{ref4}.
It is fully separable if it is a probabilistic mixture of tensor products of one-qubit states.
For the main counterexample, we use the discrete-gate formulation of $\mathrm{DQC1}_{\mathrm{sep}}$, which requires that there is no entanglement across any bipartition after each gate~\cite[Section~4.2]{ref4}.

\paragraph{Required eigenproperties.} Definition~4.2.3 of Ref.~\cite{ref4} states that an $n$-qubit unitary $U$ has a separable eigendecomposition if (a) there exists a set of separable eigenvectors $\{|u_{j}\rangle\}_{j}$ and (b) there is a probability distribution $p$ such that $\sum_{j} p_{j}|u_{j} \rangle \langle u_{j}| = \mathbb{I}/2^{n}$.
Each $n$-qubit unitary related to our counterexample has a set of $2^{n}$ eigenvectors that are fully separable and orthonormal to one another.
Such unitaries satisfy condition (b) with a uniform probability distribution.

\paragraph{Why these properties guarantee separability.} As explained in Ref.~\cite{ref4}, it is conceptually helpful to express the maximally mixed state of the register qubits as $(\mathbb{I}/2^{n})=2^{-n}\sum_{i}|u_{i}\rangle\langle u_{i}|$, where $\{|u_{i}\rangle\}$ is an orthonormal eigenbasis for $U$.
It is notationally convenient to denote the mixed state $\sum_{i} p_{i}|\psi_{i} \rangle \langle \psi_{i}|$ as $\{p_{i}, |\psi_{i}\rangle\}_{i}$, where $p$ is a probability distribution.
The initial state of the circuit is therefore $\{2^{-n}, |+\rangle |u_{i}\rangle \}_{i}$.
Let $\lambda_{i}$ be the eigenvalue of $|u_{i}\rangle$.
When $U$, controlled by the clean qubit, is applied, the state becomes
\begin{equation}\label{eq:ensemble_approach}
\frac{|0\rangle |u_{i}\rangle + |1\rangle \lambda_{i}|u_{i}\rangle}{\sqrt{2}}
= 
\frac{|0\rangle + \lambda_{i}|1\rangle}{\sqrt{2}} |u_{i}\rangle.
\end{equation}
This equation clearly has no entanglement if $\{|u_{i}\rangle\}_{i}$ is a separable eigendecomposition.
The final state before the measurement is a probabilistic mixture of these states and is therefore fully separable.

\section{The counterexample and its separability}\label{sec:counterexample}

We first describe the counterexample circuit shown in Figure~\ref{fig:counterexample} and then prove that it is fully separable after every gate.

\subsection{Quantum circuit}\label{sec:circuit}

\paragraph{The register and gates.} The register has three maximally mixed qubits, $a$, $b$ and $c$.
The clean qubit that controls the operations on these qubits is not shown.
Initially, a CNOT gate is applied with $a$ as the control and $b$ as the target.
This is followed by a gate that consists of an open-controlled $Z$ on $c$ that is controlled by $a$, and a $T$ gate on $a$.
Finally, a CNOT gate is applied with $b$ as the control and $c$ as the target.
These gates can be specified as:
\begin{equation}
\begin{aligned}
G_{1} &= \operatorname{CNOT}_{a\to b},\\
G_{2} &= |0\rangle\langle0|_{a}\otimes Z_{c}
       + e^{i \pi / 4}|1\rangle\langle1|_{a}\otimes \mathbb{I}_{c},\\
G_{3} &= \operatorname{CNOT}_{b\to c}.
\end{aligned}
\label{eq:gates}
\end{equation}

\paragraph{The accumulated unitaries.} Let $C_{t}$ denote the accumulated unitary after $t$ gates have been applied.
That is, $C_{t} = G_{t} G_{t-1} \dots G_{1}$.
Direct calculation gives:
\begin{equation}\label{eq:accumulated_unitaries}
\begin{aligned}
C_{0} &= \mathbb{I},\\
C_{1} &= \operatorname{CNOT}_{a\to b},\\
C_{2} &= |0\rangle\langle0|_{a} \otimes C_{2,0} + |1\rangle\langle1|_{a} \otimes C_{2,1},\\
C_{3} &= |0\rangle\langle0|_{a} \otimes C_{3,0} + |1\rangle\langle1|_{a} \otimes C_{3,1},
\end{aligned}
\end{equation}
where
\begin{equation}\label{eq:V0_and_V1}
\begin{aligned}
C_{2,0} &= Z_{c},\\
C_{2,1} &= e^{i \pi / 4} X_{b},\\
C_{3,0} &= |0\rangle\langle0|_{b} \otimes Z_{c} + |1\rangle \langle1|_{b} \otimes(XZ)_{c},\\
C_{3,1} &= e^{i \pi / 4} (X_{b} \otimes |+\rangle\langle+|_{c} + (ZX)_{b} \otimes |-\rangle\langle-|_{c}).
\end{aligned}
\end{equation}

\subsection{Separability after every gate}\label{sec:gate-separability}

Each of the accumulated 3-qubit unitaries, $C_{0}$ to $C_{3}$, has a set of 8 fully separable, orthonormal eigenvectors.
Direct calculation shows that the accumulated unitaries $C_{0}$ to $C_{2}$ have degenerate eigenvalues, while $C_{3}$ is non-degenerate.
One set of separable, orthonormal eigenvectors that conveniently works for $C_{0}$, $C_{1}$ and $C_{2}$ is
\begin{equation}
   |z_{\pm}\rangle |x_{\pm}\rangle |z_{\pm}\rangle,
\end{equation}
with signs chosen independently and subscripts, $a$, $b$ and $c$, read from left to right, suppressed for clarity.
Inspection of Eqs.~\eqref{eq:accumulated_unitaries} and~\eqref{eq:V0_and_V1} gives the eigenvectors of $C_{3}$ as shown in Table~\ref{tab:eigenvectors}.

\begin{table}[htbp]
\centering
\begin{tabular}{@{}lcc@{\qquad\qquad}lcc@{}}
\toprule
Eigenvector & Eigenvalue & $k$ & Eigenvector & Eigenvalue & $k$\\
\midrule
$|0\rangle |0\rangle     |0\rangle$     & $1$                 & $0$ &
$|1\rangle |x_{+}\rangle |+\rangle$     & $e^{i \pi / 4}$     & $1$\\
$|0\rangle |0\rangle     |1\rangle$     & $-1$                & $4$ &
$|1\rangle |x_{-}\rangle |+\rangle$     & $-e^{i \pi / 4}$    & $5$\\
$|0\rangle |1\rangle     |y_{+}\rangle$ & $-i$                & $6$ &
$|1\rangle |y_{+}\rangle |-\rangle$     & $i\,e^{i \pi / 4}$  & $3$\\
$|0\rangle |1\rangle     |y_{-}\rangle$ & $i$                 & $2$ &
$|1\rangle |y_{-}\rangle |-\rangle$     & $-i\,e^{i \pi / 4}$ & $7$\\
\bottomrule
\end{tabular}
\caption{Separable, orthonormal eigenvectors of $C_{3}$.
Each eigenvalue equals $e^{i k \pi / 4}$ for the listed~$k$.}
\label{tab:eigenvectors}
\end{table}

\paragraph{Continuous version.} In Appendix~\ref{app:continuous}, we give explicit unitary interpolations that connect $C_{0}, C_{1}, C_{2}$ and $C_{3}$ while keeping a separable, orthonormal eigenbasis at every intermediate time.
Each gate acts on its original pair of register qubits, with the earlier gates already completed.
The resulting evolution stays fully separable throughout and has the same final accumulated unitary, $C_{3}$.

\section{The canonical form of \texorpdfstring{$\mathrm{DQC1}_{\mathrm{sep}}$}{DQC1sep} circuits}
\label{sec:canonical}

We first use the block decomposition of a unitary to define fixed bases and their branches.
This feeds into the definition of canonical form.
For completeness, we also point the reader towards the parts of Ref.~\cite{ref4} that describe the role of canonical form in the efficient, classical simulations of $\mathrm{DQC1_{sep}}$ circuits.

\subsection{Fixed bases}
\label{sec:bases}

\paragraph{Definition.} If, for a unitary, $U$, there exists a block decomposition of the form
\begin{equation}
  U = |e_{0}\rangle \langle e_{0}|_{j} \otimes U_{0}
    + |e_{1}\rangle \langle e_{1}|_{j} \otimes U_{1},
  \label{eq:fixedbasis}
\end{equation}
where $|e_{0}\rangle$ and $|e_{1}\rangle$ form an orthonormal basis for qubit $j$, and the unitaries $U_{0}$ and $U_{1}$ act on the other qubits, then qubit $j$ is said to have a fixed basis.
If no such block decomposition exists, then qubit $j$ does not have a fixed basis.

This is equivalent to the fixed basis line of Definition~4.3.7 in Ref.~\cite{ref4}, which requires that inputting $|e_{0}\rangle$ or $|e_{1}\rangle$ on qubit $j$, with arbitrary inputs on the other qubits, returns the same basis state on qubit $j$.

\paragraph{Uniqueness.} If $U_{0} \ne U_{1}$, then qubit $j$ has a fixed basis formed by $|e_{0}\rangle$ and $|e_{1}\rangle$ that is unique up to phases and reordering.
If, on  the other hand, $U_{0} = U_{1}$, then $U = \mathbb{I}_{j} \otimes U_{0}$, so qubit $j$ has a fixed basis that is not unique (i.e., its fixed basis could be any orthonormal single-qubit basis).

\paragraph{Branching example.} It is convenient to describe $U_{0}$ and $U_{1}$ as the branches or the branch unitaries of the fixed basis.
A pertinent example is that of $C_{3}$, with its unique branching structure illustrated in Figure~\ref{fig:branches}.

\BranchStructureFigure

\subsection{Canonical form}
\label{sec:canonical-form}

Definition~4.3.10 of Ref.~\cite{ref4} states that an $n$-qubit unitary, $U$, is in canonical form if there exist $s$ wires, collectively called $B$, with fixed bases $\{ |b_{0}^{(i)}\rangle, |b_{1}^{(i)}\rangle \}$, and single-qubit unitaries $W^{j,x}$ acting on the remaining $q = n - s$ qubits, denoted $T_{1}, \dots, T_{q}$, such that
\begin{equation}
  \begin{aligned}
    U &= \sum_{x \in \{ 0,1 \}^s} e^{i \theta_x} \,|x\rangle \langle x|_{B}
         \otimes W^{1,x}_{T_{1}} \otimes \dots \otimes W^{q,x}_{T_{q}},\\
    |x\rangle &= |b^{(1)}_{x_1}\rangle \otimes \dots \otimes |b^{(s)}_{x_s}\rangle.
  \end{aligned}
  \label{eq:canonical}
\end{equation}

The equation has a natural branching interpretation (see Figure~\ref{fig:canonical_form}).
The fixed basis qubits in $B$ define a common, ordered, branching tree.
At stage $i$, the current branch splits according to the value $x_{i}$ of the $i$th control qubit in its fixed basis.
Each leaf, $x \in \{0, 1 \}^{s}$, then carries the phase $e^{i \theta_{x}}$ and the target unitary is a tensor product of single-qubit unitaries, $\bigotimes_{j=1}^{q}W^{j,x}_{T_{j}}$, acting on the remaining qubits.

\CanonicalFormStructure

A unitary \emph{can be put in canonical form} if it is possible to make a single, consistent choice about which qubits with fixed bases are members of $B$ (and which aren't), and to specify their fixed bases, phases and tensor products of single-qubit unitaries.

\subsection{Canonical form and efficient classical simulation}

\SimulationScopeFigure

Figure~\ref{fig:scope} summarizes the simulation argument of the thesis~\cite{ref4} and, anticipating Section~\ref{sec:noncanonical}, shows where our counterexample lies in relation to it.
Theorem~4.2.5 of the thesis~\cite{ref4} states that each accumulated unitary of a $\mathrm{DQC1}_{\mathrm{sep}}$ circuit must have a separable eigendecomposition.
The thesis defines a product control unitary as one implemented by a product control circuit.
These circuits are built up, gate by gate, according to specified rules so that each accumulated unitary has a separable eigendecomposition.
Theorem~4.3.9 asserts that every $\mathrm{DQC1}_{\mathrm{sep}}$ circuit is a product control circuit.

Taken together, the aforementioned theorems claim that product control circuits are the only type of $\mathrm{DQC1}_{\mathrm{sep}}$ circuit (see Figure~\ref{fig:scope}(a)).
In other words, product control circuits exhaustively characterize $\mathrm{DQC1}_{\mathrm{sep}}$ circuits.
Theorems~4.4.4 and~4.4.5 are used in the proof of Theorem~4.3.9, and their proofs both contain the necessity claim (see Section~\ref{sec:false-necessity}) that our counterexample shows is false.

The proof of Theorem~4.3.11 of the thesis~\cite{ref4} contains an efficient classical algorithm that takes a polynomial-size description of a product control circuit as input and outputs a compact representation of its final accumulated unitary in canonical form:
the fixed bases, together with a way of efficiently computing, for any control string $x\in\{0,1\}^{s}$, the phase $e^{i \theta_{x}}$ and the single-qubit unitaries $W^{j,x}$ of Equation~\eqref{eq:canonical}.

The proof of Theorem~4.3.12 contains an efficient classical algorithm that takes this compact representation as input, uniformly samples the (possibly degenerate) eigenvalues of the final accumulated unitary, and thereby estimates the normalized trace to inverse-polynomial additive accuracy (see Figure~\ref{fig:scope}(b)).
All together, this decides any problem in $\mathrm{DQC1}_{\mathrm{sep}}$, which is how the thesis concludes that $\mathrm{DQC1}_{\mathrm{sep}}\subseteq\mathrm{BPP}$.

\section{Proof that the counterexample cannot be put into canonical form}
\label{sec:noncanonical}

We first determine the fixed-basis qubits of the final accumulated unitary, $C_{3}$, of our counterexample circuit. 
We analyze the possible branching structures to rule out every canonical representation.
The correctness of this analysis implies that there must be a gap in the proof of Ref.~\cite{ref4}.
We identify this as a false necessity condition, and describe how our counterexample evades it.

\subsection{Fixed bases and noncanonicality}
\label{sec:fixed-bases}

\begin{theorem}[Failure of canonical form]
\label{theorem:noncanonical}
For the final unitary, $C_{3}$, of the counterexample circuit:
\begin{enumerate}
  \item the $Z$ basis is the only fixed basis of qubit $a$, up to phases and reordering;
  \item neither qubit $b$ nor qubit $c$ has a fixed basis;
  \item $C_{3}$ cannot be put in canonical form.
\end{enumerate}
\end{theorem}

\begin{proof}

\noindent \textbf{Unique fixed basis for $a$.} By Equation~\eqref{eq:accumulated_unitaries}, qubit $a$ has a fixed basis, namely the $Z$ basis.
Since the two branch unitaries, $C_{3,0}$ and $C_{3,1}$, of Equation~\eqref{eq:V0_and_V1} are unequal, this basis is unique, up to phases and reordering, which establishes part 1 of the theorem.
\smallskip

\noindent \textbf{No fixed basis for $b$ or $c$.} Suppose that qubit $b$ had a fixed basis $\{|e_{0}\rangle, |e_{1}\rangle\}$.
Then Equation~\eqref{eq:fixedbasis} would imply that every eigenvector of $C_{3}$ has qubit $b$ in $|e_{0}\rangle$ or $|e_{1}\rangle$.
Because $C_{3}$ is non-degenerate, its eigenbasis is unique up to phases and reordering.
This eigenbasis is shown in Table~\ref{tab:eigenvectors}, where $b$ appears in six different states, $|x_{\pm}\rangle$, $|y_{\pm}\rangle$ and $|z_{\pm}\rangle$, rather than in a single orthonormal pair, as is required.
Hence, $C_{3}$ cannot be put into a form in which $b$ has a fixed basis.
The same argument applies to qubit $c$. 
This establishes part 2 of the theorem.
\smallskip

\noindent \textbf{Candidates for \textit{B}.} Qubit $a$ is the only qubit with a fixed basis, so either $B=\varnothing$ or $B=\{a\}$ (i.e., either $B$ is empty or it contains qubit $a$).
\smallskip

\noindent \textbf{Canonical form.} If $B=\varnothing$, then a canonical representation (Equation~\eqref{eq:canonical}) would require that $C_{3} = e^{i \theta} (W_{a} \otimes W_{b} \otimes W_{c})$.
If $B=\{ a \}$, then it would require that $C_{3} = |0\rangle \langle0|_{a} \otimes e^{i \theta_{0}} (W_{b}^{0} \otimes W_{c}^{0}) + |1\rangle \langle1|_{a} \otimes e^{i \theta_{1}} (W_{b}^{1} \otimes W_{c}^{1})$.
\smallskip

\noindent \textbf{Implications for the block decomposition.} Let us consider the $|0\rangle \langle0|_{a}$ block of these updated accumulated unitaries.
In the former case, it is of the form $e^{i \theta} \, \mathrm{}_{a}\langle0| W_{a}|0\rangle_{a} \, (W_{b} \otimes W_{c})$.
In the latter case, it is simply $e^{i \theta_{0}} \, (W_{b}^{0} \otimes W_{c}^{0})$.

Let us proceed to consider the $|0\rangle \langle0|_{a} \otimes |0\rangle \langle0|_{b}$ and the $|0\rangle \langle0|_{a} \otimes |1\rangle \langle1|_{b}$ blocks.
In the former case, they are of the forms $e^{i \theta} \, \mathrm{}_{a}\langle0| W_{a}|0\rangle_{a} \, \mathrm{}_{b}\langle0|W_{b}|0\rangle_{b} \, W_{c}$ and $e^{i \theta} \, \mathrm{}_{a}\langle0| W_{a}|0\rangle_{a} \, \mathrm{}_{b}\langle1|W_{b}|1\rangle_{b} \, W_{c}$, respectively. 
Both are scalar multiples of $W_{c}$, so they are linearly dependent.

In the latter case, they are of the forms $e^{i \theta_{0}} \, \mathrm{}_{b}\langle0|W_{b}^{0}|0\rangle_{b} \, W_{c}^{0}$ and $e^{i \theta_{0}} \, \mathrm{}_{b}\langle1|W_{b}^{0}|1\rangle_{b} \, W_{c}^{0}$, respectively.
Both are scalar multiples of $W_{c}^{0}$, so they are linearly dependent.
\smallskip

\noindent \textbf{Finding a contradiction.} According to Equations~\eqref{eq:accumulated_unitaries} and \eqref{eq:V0_and_V1}, the $|0\rangle \langle0|_{a} \otimes |0\rangle \langle0|_{b}$ block of $C_{3}$ is $Z_{c}$, and the $|0\rangle \langle0|_{a} \otimes |1\rangle \langle1|_{b}$ block is $(XZ)_{c}$.
Since the former is diagonal and the latter is off-diagonal, they are clearly linearly independent.
This is a contradiction with the block decomposition implied by the required canonical form.
Hence, $C_{3}$ cannot be put into canonical form, which establishes part 3 of the theorem.

\end{proof}

\subsection{The false necessity condition in the characterization argument}
\label{sec:false-necessity}

After $G_2$ has been applied, $b$ has only experienced $\mathbb{I}$ or $X$, and $c$ has only experienced $Z$ or $\mathbb{I}$, depending on $a$.
Those alternatives are diagonal in fixed bases but depend nontrivially on another qubit.
The thesis~\cite{ref4} calls qubits such as $b$ and $c$ `ambiguous.'
Its Theorem~4.4.4 claims to constrain the next two-qubit gate between two such lines, which, in our case, is $G_{3}$~\cite{comment}.

Recall that the branch unitaries of $C_{2}$ are $C_{2,0} = \mathbb{I}_{b} \otimes Z_{c}$ and $C_{2,1} = e^{i \pi / 4} (X_{b} \otimes \mathbb{I}_{c})$, so the updated branches are $C_{3,x} = G_{3} C_{2,x}$ for $x \in \{0,1\}$.
The proof starts by requiring that in each updated branch (in our notation, $C_{3,x}$), at least one qubit ($b$ or $c$ or both) must have a fixed basis.
Importantly, each updated branch can satisfy this requirement \emph{in its own way}.

The proof proceeds to suppose that the original branches can be written as $C_{2,0} = K Q_{0}$ and $C_{2,1} = K Q_{1}$, where $Q_{0}$ and $Q_{1}$ are both diagonal in some common product basis.
The idea is that since $K$ is a general two-qubit unitary, this product basis does not have to coincide with either the $X$ basis of $b$ or the $Z$ basis of $c$.
Substituting this equation for the original branches into the one for the updated branches gives $C_{3,x} = G_{3}(K Q_{x}) = (G_{3} K) Q_{x}$. 

From here, the proof claims that $G_{3} K$ must have at least one fixed basis (on $b$ or $c$ or both) that coincides with the bases in which $Q_{0}$ and $Q_{1}$ are diagonal.
However, $G_{3} K$ has no branch-dependence.
Consequently, this requirement implies that the updated branches satisfy this requirement \emph{in the same way} (namely, that the same qubit has the same fixed basis in both updated branches).
Although satisfying the requirement in the same way is sufficient, it is not necessary (see Sections~\ref{sec:gate-separability} and~\ref{sec:fixed-bases}).

\subsection{How the counterexample evades the claimed necessity}
\label{sec:evades}

Building on the previous subsection, we now describe how the counterexample circuit evades the claimed necessary condition.
If $Q_{0}$ and $Q_{1}$ are diagonal in a product basis, then $Q_{0}^{\dagger} Q_{1}$ must also be diagonal in that basis.
We can find this basis by evaluating $C_{2,0}^{\dagger} C_{2,1} = (K Q_{0})^{\dagger} (K Q_{1}) = Q_{0}^{\dagger} Q_{1} = e^{i \pi / 4} (X_{b} \otimes Z_{c})$, which despite being degenerate necessitates that any product eigenvector must have an $X$ eigenvector on $b$ and a $Z$ eigenvector on $c$.
Although this basis did not have to coincide with either the $X$ basis of $b$ or the $Z$ basis of $c$, for our counterexample circuit, it did.

We can now test whether $G_{3} K$ can be controlled in this basis by seeing if it commutes with $X_{b} \otimes I_{c}$ or $I_{b} \otimes Z_{c}$.
Since $C_{2,0}, C_{2,1}, Q_{0}$ and $Q_{1}$ are all diagonal in the product of the $X$ basis on $b$ and the $Z$ basis on $c$, and $C_{2,x}  = K Q_{x}$, we know that $K$ is also diagonal in this basis.
Consequently, $[G_{3} K, X_{b} \otimes \mathbb{I}_{c}] = [G_{3}, X_{b} \otimes \mathbb{I}_{c}] \, K$ and $[G_{3} K, \mathbb{I}_{b} \otimes Z_{c}] = [G_{3},\mathbb{I}_{b} \otimes Z_{c}] \, K$.
Since $K$ is unitary, $G_{3} K$ commutes with these operators if and only if $G_{3}$ does.

Recall that $G_{3}$ is a CNOT gate with $b$ as the control and $c$ as the target.
Direct calculation shows that $G_{3}$ commutes with neither $X_{b} \otimes \mathbb{I}_{c}$ nor $\mathbb{I}_{b} \otimes Z_{c}$.
Since $C_{3}$ nevertheless has a separable eigendecomposition (see Section~\ref{sec:gate-separability}), this contradicts the necessity conditions asserted in the proof of Theorem~4.4.4 and Case~1 of the proof of Theorem~4.4.5. 
Our counterexample is therefore not captured by the forms claimed to be exhaustive.
The omitted possibility is that the updated branches are basis-controlled in different directions: $C_{3,0}$ in the $Z$ basis of $b$ and $C_{3,1}$ in the $X$ basis of $c$ (see Figure~\ref{fig:branches}). 

\section{Uniform families and classical simulation}
\label{sec:uniform}

We first extend the counterexample to a uniform family and prove that every padded member has a noncanonical final register unitary.
We then show that it is efficient to classically simulate this family.
Finally, we discuss what a repaired simulation argument would need
to establish.

\subsection{A uniform family of counterexamples}
\label{sec:uniform-family}

Complexity-theoretic statements such as $\mathrm{DQC1}_{\mathrm{sep}}\subseteq\mathrm{BPP}$ concern families of circuits, one for each number $n$ of register qubits, with $n+1$ qubits in each complete DQC1 circuit.
A family is \emph{uniform} if a classical algorithm can output a description of its member indexed by $n$ in time polynomial in $n$.
The characterization and canonicalization claims of Ref.~\cite{ref4} are stated for uniform families, so we embed the counterexample in one.

For each $n\ge3$, apply the three gates of Section~\ref{sec:circuit} to the first three register qubits and leave the other wires untouched.
The resulting accumulated unitaries are
\begin{equation}
    C_{t}^{(n)} = C_{t} \otimes \mathbb{I}^{\otimes(n-3)},\qquad t = 0, 1, 2, 3.
\end{equation}
Every member is described by the same three gates, so its description can be written down in time polynomial in $n$, and the family is uniform.
Putting the eigenbases shown in Section~\ref{sec:gate-separability} into a tensor product with the computational basis on the remaining qubits shows that every accumulated unitary has a set of $2^{n}$ fully separable, orthonormal eigenvectors.
Hence every circuit in the family is a $\mathrm{DQC1}_{\mathrm{sep}}$ circuit.

\begin{theorem}[The padded family cannot be put into canonical form]
\label{theorem:padded}
For every $n\ge3$, the accumulated unitary $C_3^{(n)} = C_3 \otimes \mathbb{I}^{\otimes(n-3)}$ cannot be put in canonical form.
\end{theorem}

\begin{proof}

For $n>3$, consider one spectator qubit, $l$.
Assume, for the sake of contradiction, that $C_{3}^{(n)} = C_{3}^{(n-1)} \otimes \mathbb{I}_{l}$ has a canonical representation (see Equation~\eqref{eq:canonical}).
There are two possibilities, either $l$ is in $C_{3}^{(n)}$'s set $B$ or it isn't, and for each of these, we need to ensure that $C_{3}^{(n-1)}$ can be put into canonical form.

If $l \in B$, we can let the first branch of $C_{3}^{(n)}$ correspond to the two vectors of qubit $l$'s fixed basis, which we could denote $|e_{0}\rangle_{l}$ and $|e_{1}\rangle_{l}$.
Since $C_{3}^{(n)} = C_{3}^{(n-1)} \otimes \mathbb{I}_{l} = |e_{0}\rangle \langle e_{0}|_{l} \otimes C_3^{(n-1)} + |e_{1}\rangle \langle e_{1}|_{l} \otimes C_{3}^{(n-1)}$, either branch gives a canonical representation of $C_{3}^{(n-1)}$ with $l$ removed from set $B$.

If $l \notin B$, then $l$ is a target qubit.
Since each branch acts trivially on $l$, every leaf must have a factor $e^{i \phi_{x}} \mathbb{I}_{l}$ for some phase $e^{i \phi_{x}}$.
Absorbing $e^{i \phi_{x}}$ into the existing branch phase and removing $\mathbb{I}_{l}$ from every leaf gives a canonical representation of $C_{3}^{(n-1)}$ with the same set $B$.

Since a canonical representation for $C_{3}^{(n)}$ would yield one for $C_{3}^{(n-1)}$ for all $n>3$, one can arrive at a canonical representation for the original counterexample accumulated unitary, $C_{3}$.
This would contradict Theorem~\ref{theorem:noncanonical}.
Therefore, no member of the padded family can be put in canonical form.

\end{proof}

\subsection{Efficient simulation of the counterexample}
\label{sec:simulation}

The counterexample of Section~\ref{sec:counterexample} has a normalized trace equal to zero.
Furthermore, padding preserves the normalized trace:
\begin{equation}
  \frac{\operatorname{Tr}[C_{3}^{(n)}]}{2^{n}}
  =\frac{2^{n-3}\operatorname{Tr}[C_{3}]}{2^{n}}
  =0.
  \label{eq:paddedtrace}
\end{equation}
Consequently, the output distribution of the counterexample circuit can be efficiently classically simulated by a fair coin.

\subsection{What a repaired simulation argument would require}
\label{sec:repair}

A repair based on tracking controlled branches must accommodate the possibility identified in Section~\ref{sec:false-necessity}, namely that the control direction may be different in different branches.
The nested description in Figure~\ref{fig:branches} illustrates this additional freedom, but the example does not establish that such descriptions suffice for every permitted circuit.
A general structural repair must cover the separability promise of Section~\ref{sec:separability} at every prefix and prove that the information used by its classical simulator can be obtained from the input circuit with polynomial resources.
If it updates that information gate by gate, its representation and update rules must remain valid through every allowed transition, including changes of control direction.

\section{Conclusion and outlook}
\label{sec:conclusion}

Aaronson asked whether a quantum computer that is always in a separable mixed state can be simulated by a classical computer in polynomial time~\cite{ref1}.
Ref.~\cite{ref3}, refined in Ref.~\cite{ref4}, argued for an affirmative answer for the one clean qubit model.
However, our  counterexample is fully separable after every gate even though its final unitary, $C_{3}$, cannot be put into the canonical form required by that argument.
We build on this by pinpointing the gap in this proof.
Nevertheless, the properties of our counterexample circuits and the insights gained from this work have not let us answer Aaronson's question one way or another.

By reopening the one clean qubit case of Aaronson's semi-grand challenge, we hope to spur further research in this area.
Our counterexample had a key property: different branches were controlled in different directions.
Theoretically, it would be interesting to study whether this property could be tracked efficiently.
Experimentally, the neutral atom systems that were once proposed as promising platforms for realizing one clean qubit protocols~\cite{ref25} have made substantial progress toward scalable fault-tolerant quantum computing~\cite{AQCE, LeRegent2025}.
Models of quantum computation related to the one clean qubit model have even been proposed as interfaces to cloud-accessed quantum computers that could limit their cryptanalytic power while still enabling useful scientific applications~\cite{ref24}.
We therefore expect that future work, both theoretical and experimental, will be of considerable interest to quantum computing researchers.

\section*{Acknowledgments}
I would like to acknowledge Leo Ouchi, Mario Duran-Vega, Cuauhtémoc Daniel Suárez Ramírez, Silvia Bergamini and Aleksei Kodukhov for helpful discussions.
I acknowledge the use of OpenAI models, especially GPT-6 Astra, and Anthropic's Claude Opus for assistance, through extensive dialogue, with refining the mathematical arguments, as well as drafting and revising the text and figures. 
I take full responsibility for the manuscript.

\section*{Dedication}
I dedicate this work to my late father.

\phantomsection

\appendix
\section{Further background on the one clean qubit model} \label{app:background}

This appendix derives the readout formulas of the one clean qubit protocol for normalized trace estimation (see Figure~\ref{fig:trace} for the circuit diagram).
In the computational basis of the clean qubit, the state of the system after the controlled unitary is
\begin{equation}
 \rho=\frac{1}{2^{n+1}}
 \begin{pmatrix} 
    \mathbb{I} & U^{\dagger} \\ 
    U          & \mathbb{I} 
 \end{pmatrix}.
 \label{eq:rho}
\end{equation}
This is the same state as that of the ensemble described in Equation~\eqref{eq:ensemble_approach}.
Tracing out the register gives the reduced density matrix of the clean qubit
\begin{equation}
 \rho_{clean}=\frac{1}{2}
 \begin{pmatrix}
  1                           & \operatorname{Tr}(U^{\dagger})/2^{n}\\
  \operatorname{Tr}(U)/2^{n}  &1
 \end{pmatrix}.
 \label{eq:reduced}
\end{equation}
The expectation values of Pauli measurements of the clean qubit are therefore
\begin{equation}
 \langle X \rangle = \frac{\operatorname{Re}[\operatorname{Tr}(U)]}{2^{n}},
 \qquad
 \langle Y \rangle = \frac{\operatorname{Im}[\operatorname{Tr}(U)]}{2^{n}}.
 \label{eq:expectations}
\end{equation}

Equivalently, the probability of the $\pm1$ outcome in an $X$ measurement is $\frac{1}{2}(1\pm\operatorname{Re}[\operatorname{Tr}(U)]/2^{n})$.
For $Y$, it is $\frac{1}{2}(1\pm\operatorname{Im}[\operatorname{Tr}(U)]/2^{n})$.
Repeating either measurement polynomially many times estimates the corresponding expectation value to inverse-polynomial additive accuracy with bounded failure probability~\cite[Section~4.1]{ref4}.

\section{Continuous counterexample} \label{app:continuous}

\begin{continuousproposition}[Continuous separability]

There is a continuous, piecewise smooth register-unitary path $C(s)$, $0\leq s\leq3$, with $C(0)=C_{0}$, $C(1)=C_{1}$, $C(2)=C_{2}$ and $C(3)=C_{3}$, such that $C(s)$ has a separable, orthonormal eigenbasis at every time.
Consequently, the corresponding clean-controlled implementation remains fully separable throughout and ends at the noncanonical, final unitary, $C_{3}$, of Theorem~\ref{theorem:noncanonical}.

\end{continuousproposition}

\begin{proof}

\textbf{Interpolating the gates.} For $P \in \{ X,Z \}$ and $0 \leq t \leq 1$, define
\begin{equation}
 R_{P}(t)=\frac{\mathbb{I} + P}{2}
       +e^{i \pi t}  \, \frac{\mathbb{I} - P}{2}.
 \label{eq:continuous-rp}
\end{equation}
This enables the following continuous gates to be defined.
\begin{equation}
\begin{aligned}
G_{1}(t) &= |0\rangle\langle0|_{a} \otimes \mathbb{I}_{b}        + |1\rangle \langle1|_{a} \otimes R_{X_{b}}(t),\\
G_{2}(t) &= |0\rangle\langle0|_{a} \otimes R_{Z_{c}}(t) + e^{i \pi t / 4} \, |1\rangle \langle1|_{a} \otimes \mathbb{I}_{c},\\
G_{3}(t) &= |0\rangle\langle0|_{b} \otimes \mathbb{I}_{c}        + |1\rangle \langle1|_{b} \otimes R_{X_{c}}(t).
\end{aligned}
\label{eq:continuous_gates}
\end{equation}
The paths begin at the identity and end at $G_{1}$, $G_{2}$ and $G_{3}$, respectively.
Applying them sequentially gives the accumulated unitary
\begin{equation}
 C(s)=
 \begin{cases}
  G_{1}(s), & 0\leq s\leq1,\\
  G_{2}(s-1) G_{1}, & 1< s\leq2,\\
  G_{3}(s-2) G_{2} G_{1}, & 2< s\leq3,
 \end{cases}
 \label{eq:continuous-path}
\end{equation}
which, recalling Equation~\eqref{eq:accumulated_unitaries}, agrees with the accumulated unitaries of the discrete counterexample at integer values of~$s$.

During the first two stages (i.e., $0 \leq s \leq 2$), $C(s)$ is diagonal in the tensor product of the $Z$ basis on $a$, the $X$ basis on $b$ and the $Z$ basis on $c$. The accumulated unitary during the final stage (i.e., $2 < s \leq 3$) can be expressed as
\begin{equation}
C(s) = |0\rangle \langle0|_{a} \otimes C_{3,0}(s) + |1\rangle \langle1|_{a} \otimes C_{3,1}(s),
\end{equation}
where
\begin{equation}\label{eq:continuousV0_and_V1}
\begin{aligned}
C_{3,0}(s) &= |0\rangle \langle0|_{b} \otimes Z_{c} + |1\rangle \langle1|_{b} \otimes (R_{X_{c}}(s-2)Z_{c}),\\
C_{3,1}(s) &= e^{i \pi / 4} (X_{b} \otimes |+\rangle \langle+|_{c} + (R_{Z_{b}}(s-2)X_{b}) \otimes |-\rangle \langle-|_{c}).
\end{aligned}
\end{equation}

The eigenvectors of $C(s)$, for $2 < s \leq 3$, are shown in Table~\ref{tab:eigenvectors_s}.
They form a separable, orthonormal eigenbasis.
Crucially, $C(s=3) = C_{3}$ and by Theorem~\ref{theorem:noncanonical}, $C_{3}$ cannot be put into canonical form.

\begin{table}[htbp]
\centering
\setlength{\aboverulesep}{0pt}
\setlength{\belowrulesep}{0pt}
\renewcommand{\arraystretch}{1.4}
\begin{tabular}{@{}lc@{\qquad}|@{\qquad}lc@{}}
\toprule
\multicolumn{2}{@{}c@{\qquad}|@{\qquad}}{$a=0$ branch} &
\multicolumn{2}{c@{}}{$a=1$ branch}\\
\cmidrule(r){1-2}\cmidrule(l){3-4}
Eigenvector & Eigenvalue & Eigenvector & Eigenvalue\\
\midrule
$|0\rangle |0\rangle             |0\rangle$                    & $1$ &
$|1\rangle |x_{+}\rangle         |+\rangle$                    & $e^{i \pi / 4}$\\
$|0\rangle |0\rangle             |1\rangle$                    & $-1$ &
$|1\rangle |x_{-}\rangle         |+\rangle$                    & $-e^{i \pi / 4}$\\
$|0\rangle |1\rangle             |\tilde{z}_{-}\rangle$        & $-e^{i \pi t / 2}$ &
$|1\rangle |\tilde{x}_{+}\rangle |-\rangle$                    & $e^{i \pi t / 2} \, e^{i \pi / 4}$\\
$|0\rangle |1\rangle             |\tilde{z}_{+}\rangle$        & $e^{i \pi t / 2}$ &
$|1\rangle |\tilde{x}_{-}\rangle |-\rangle$                    & $-e^{i \pi t / 2} \, e^{i \pi / 4}$\\
\bottomrule
\end{tabular}
\caption{A separable, orthonormal eigenbasis of $C(s)$ for $2 < s \le 3$, with factors ordered as $a,b,c$,
where $t = s - 2$, $|\tilde{z}_{\pm}\rangle = e^{-i \pi t X / 4} \, |z_{\pm}\rangle$ and $|\tilde{x}_{\pm}\rangle = e^{-i \pi t Z / 4} \, |x_{\pm}\rangle$.
At $s=3$, $|\tilde{z}_{\pm}\rangle \propto |y_{\mp}\rangle$ and $|\tilde{x}_{\pm}\rangle \propto |y_{\pm}\rangle$, so each row reduces, up to a phase, to the corresponding row of Table~\ref{tab:eigenvectors}.
The eigenvalues are distinct except at $s=5/2$, where $e^{i \pi t / 2} = e^{i \pi / 4}$ and the last two rows of the $a=0$ branch become degenerate with the first two rows of the $a=1$ branch.}
\label{tab:eigenvectors_s}
\end{table}

\end{proof}

Following Section~\ref{sec:uniform} on uniform families, let $C^{(n)}(s)=C(s) \otimes \mathbb{I}^{\otimes(n-3)}$.
Putting the eigenbases above into a tensor product with the computational basis states on the spectator wires proves full separability throughout every padded implementation.


\addcontentsline{toc}{section}{References}
\begin{thebibliography}{27}

\bibitem{ref1}
S.~Aaronson, ``Ten semi-grand challenges for quantum computing theory'' (2005),
\url{https://www.scottaaronson.com/writings/qchallenge.html}.

\bibitem{ref2}
E.~Knill and R.~Laflamme, ``Power of one bit of quantum information,'' \emph{Phys. Rev. Lett.} \textbf{81},
5672--5675 (1998). \href{https://doi.org/10.1103/PhysRevLett.81.5672}{doi:10.1103/PhysRevLett.81.5672}.

\bibitem{ref100}
L.~Susskind, \emph{Three Lectures on Complexity and Black Holes}, SpringerBriefs in Physics (Springer Cham, 2020).
\href{https://doi.org/10.1007/978-3-030-45109-7}{doi:10.1007/978-3-030-45109-7}.

\bibitem{ref101}
D.~J. Shepherd, ``Computation with unitaries and one pure qubit,''
\href{https://arxiv.org/abs/quant-ph/0608132}{arXiv:quant-ph/0608132} (2006).

\bibitem{ref11}
K.~Fujii, H.~Kobayashi, T.~Morimae, H.~Nishimura, S.~Tamate, and S.~Tani, ``Impossibility of classically simulating one-clean-qubit model with multiplicative error,'' \emph{Phys. Rev. Lett.}
\textbf{120}, 200502 (2018). \href{https://doi.org/10.1103/PhysRevLett.120.200502}{doi:10.1103/PhysRevLett.120.200502}.

\bibitem{ref12}
T.~Morimae, K.~Fujii, and J.~F.~Fitzsimons, ``Hardness of classically simulating the one-clean-qubit model,'' \emph{Phys. Rev. Lett.} \textbf{112}, 130502 (2014).
\href{https://doi.org/10.1103/PhysRevLett.112.130502}{doi:10.1103/PhysRevLett.112.130502}.

\bibitem{ref13}
T.~Morimae, ``Hardness of classically sampling the one-clean-qubit model with constant total variation distance error,'' \emph{Phys. Rev. A} \textbf{96}, 040302(R) (2017). 
\href{https://doi.org/10.1103/PhysRevA.96.040302}{doi:10.1103/PhysRevA.96.040302}.

\bibitem{ref5}
P.~W.~Shor and S.~P.~Jordan, ``Estimating Jones polynomials is a complete problem for one clean qubit,'' \emph{Quantum Inf. Comput.} \textbf{8}, 681--714 (2008), \href{https://doi.org/10.26421/QIC8.8-9-1}{doi:10.26421/QIC8.8-9-1}.

\bibitem{JordanWocjan2009} S.~P.~Jordan and P.~Wocjan, ``Estimating Jones and HOMFLY polynomials with one clean qubit,'' \emph{Quantum Inf. Comput.} \textbf{9}, 264--289 (2009), \href{https://doi.org/10.26421/QIC9.3-4-6}{doi:10.26421/QIC9.3-4-6}.

\bibitem{JordanAlagic2014} S.~P.~Jordan and G.~Alagic, ``Approximating the Turaev--Viro invariant of mapping tori is complete for one clean qubit,'' in \emph{TQC 2011}, LNCS \textbf{6745}, 53--72 (Springer, 2014), \href{https://doi.org/10.1007/978-3-642-54429-3_5}{doi:10.1007/978-3-642-54429-3\_5}.

\bibitem{ref7}
A.~N.~Chowdhury, R.~D.~Somma, and Y.~Suba\c{s}\i{}, ``Computing partition functions in the one-clean-qubit model,'' \emph{Phys. Rev. A} \textbf{103}, 032422 (2021).
\href{https://doi.org/10.1103/PhysRevA.103.032422}{doi:10.1103/PhysRevA.103.032422}.

\bibitem{ref8}
S.~Roy Moulik and S.~Strelchuk, ``DQC1-hardness of estimating correlation functions,''
\href{https://arxiv.org/abs/2411.05208}{arXiv:2411.05208} (2024).

\bibitem{LoweKBH2026} D.~Lowe, M.~S.~Kim, R.~Bondesan, and R.~Hayakawa, ``Complexity of normalized persistence problems for topological data analysis and local Hamiltonians,'' \href{https://arxiv.org/abs/2607.03278}{arXiv:2607.03278} (2026).

\bibitem{GCD2022} C.~Gyurik, C.~Cade, and V.~Dunjko, ``Towards quantum advantage via topological data analysis,'' \emph{Quantum} \textbf{6}, 855 (2022), \href{https://doi.org/10.22331/q-2022-11-10-855}{doi:10.22331/q-2022-11-10-855}.

\bibitem{CadeCrichigno2024} C.~Cade and P.~M.~Crichigno, ``Complexity of supersymmetric systems and the cohomology problem,'' \emph{Quantum} \textbf{8}, 1325 (2024), \href{https://doi.org/10.22331/q-2024-04-30-1325}{doi:10.22331/q-2024-04-30-1325}.

\bibitem{ref6}
C.~Cade and A.~Montanaro, ``The quantum complexity of computing Schatten $p$-norms,'' in \emph{13th Conference on the Theory of Quantum Computation, Communication and Cryptography (TQC 2018)}, LIPIcs \textbf{111}, 4:1--4:20 (2018). \href{https://doi.org/10.4230/LIPIcs.TQC.2018.4}{doi:10.4230/LIPIcs.TQC.2018.4}.

\bibitem{Khatri2019} S.~Khatri, R.~LaRose, A.~Poremba, L.~Cincio, A.~T.~Sornborger, and P.~J.~Coles, ``Quantum-assisted quantum compiling,'' \emph{Quantum} \textbf{3}, 140 (2019), \href{https://doi.org/10.22331/q-2019-05-13-140}{doi:10.22331/q-2019-05-13-140}.

\bibitem{ref24}
S.~A.~Yadavalli, J.~Rajakumar, A.~Schuckert, and M.~J.~Gullans, ``Protecting Quantum Computers against Untrusted Users,''
\href{https://arxiv.org/abs/2610.06812v1}{arXiv:2610.06812v1} (2026).

\bibitem{ref10}
T.~Morimae, K.~Fujii, and H.~Nishimura, ``Power of one nonclean qubit,'' \emph{Phys. Rev. A} \textbf{95},
042336 (2017). \href{https://doi.org/10.1103/PhysRevA.95.042336}{doi:10.1103/PhysRevA.95.042336}.

\bibitem{JacobsMehraban2024} D.~Jacobs and S.~Mehraban, ``The space just above one clean qubit,'' \href{https://arxiv.org/abs/2410.08051}{arXiv:2410.08051} (2024).

\bibitem{DFC2005} A.~Datta, S.~T.~Flammia, and C.~M.~Caves, ``Entanglement and the power of one qubit,'' \emph{Phys. Rev. A} \textbf{72}, 042316 (2005), \href{https://doi.org/10.1103/PhysRevA.72.042316}{doi:10.1103/PhysRevA.72.042316}.

\bibitem{ref17}
A.~Datta, A.~Shaji, and C.~M.~Caves, ``Quantum discord and the power of one qubit,'' \emph{Phys. Rev. Lett.} \textbf{100}, 050502 (2008). \href{https://doi.org/10.1103/PhysRevLett.100.050502}{doi:10.1103/PhysRevLett.100.050502}.

\bibitem{Matera2016}
J.~M.~Matera, D.~Egloff, N.~Killoran, and M.~B.~Plenio, ``Coherent control of quantum systems as a resource theory,'' \emph{Quantum Sci. Technol.} \textbf{1}, 01LT01 (2016).
\href{https://doi.org/10.1088/2058-9565/1/1/01LT01}{doi:10.1088/2058-9565/1/1/01LT01}.

\bibitem{Wang2019}
W.~Wang, J.~Han, B.~Yadin, Y.~Ma, J.~Ma, W.~Cai, Y.~Xu, L.~Hu, H.~Wang, Y.~P.~Song, M.~Gu, and L.~Sun, ``Witnessing quantum resource conversion within deterministic quantum computation using one pure superconducting qubit,'' \emph{Phys. Rev. Lett.} \textbf{123}, 220501 (2019).
\href{https://doi.org/10.1103/PhysRevLett.123.220501}{doi:10.1103/PhysRevLett.123.220501}.

\bibitem{ref26}
S.~J.~Evered, D.~Bluvstein, M.~Kalinowski, S.~Ebadi, T.~Manovitz, H.~Zhou, S.~H.~Li, A.~A.~Geim, T.~T.~Wang, N.~Maskara, H.~Levine, G.~Semeghini, M.~Greiner, V.~Vuleti\'c, and M.~D.~Lukin, ``High-fidelity parallel entangling gates on a neutral-atom quantum computer,'' \emph{Nature} \textbf{622}, 268--272 (2023).
\href{https://doi.org/10.1038/s41586-023-06481-y}{doi:10.1038/s41586-023-06481-y}.

\bibitem{ref25}
C.~W.~Mansell and S.~Bergamini, ``A cold-atoms based processor for deterministic quantum computation with one qubit in intractably large Hilbert spaces,'' \emph{New J. Phys.}
\textbf{16}, 053045 (2014).
\href{https://doi.org/10.1088/1367-2630/16/5/053045}{doi:10.1088/1367-2630/16/5/053045}.

\bibitem{ref3}
M.~Yoganathan and C.~Cade, ``The one clean qubit model without entanglement is classically simulable,'' \href{https://arxiv.org/abs/1907.08224v1}{arXiv:1907.08224v1} (2019).

\bibitem{ref4}
M.~Yoganathan, \emph{The power of restricted quantum computational models}, PhD thesis, University of Cambridge (2021), Chapter~4, ``Entanglement and the quantum computing advantage.'' \href{https://doi.org/10.17863/CAM.63085}{doi:10.17863/CAM.63085}.

\bibitem{comment}
In Ref.~\cite{ref4}, Case~1 of the proof of Theorem~4.4.5 applies to two ambiguous qubits on which no preceding gates have jointly acted, so it also applies to our counterexample circuit. 
Its proof relies on a factorization of the preceding branch unitary that is a special case of the factorization considered in Theorem~4.4.4. 
The argument in Sections~\ref{sec:false-necessity}~and~\ref{sec:evades} therefore applies to both proofs.

\bibitem{AQCE}
F.-M.~Le~R\'egent, ``Awesome Quantum Computing Experiments'' (2026),
\url{https://francoismarieleregent.xyz/awesome-quantum-computing-experiments/},
accessed 17 October 2026.

\bibitem{LeRegent2025}
F.-M.~Le~R\'egent, ``Awesome Quantum Computing Experiments: Benchmarking Experimental Progress Towards Fault-Tolerant Quantum Computation,''
\href{https://arxiv.org/abs/2507.03678}{arXiv:2507.03678} (2025).

\end{thebibliography}
\end{document}